\documentclass[twocolumn,nofootinbib,showpacs,prd]{revtex4}%
\usepackage{amsmath}
\usepackage{amsfonts}
\usepackage{amssymb}
\usepackage{graphicx}

\begin{document}

\title{Entangled states in quantum cosmology and the interpretation of $\Lambda$}

\author{
Salvatore Capozziello$^\P$ and Orlando Luongo${}^\S{}^\ddag$}

\address{
$^\P$ Dipartimento di Scienze Fisiche, Universit\`a di Napoli
''Federico II'', Compl. Univ. di Monte S. Angelo, Edificio G, Via
Cinthia, I - 80126 - Napoli, Italy.}

\address{
  $^\ddag$
  Dipartimento di Fisica,
  Universit\`a di Roma ''La Sapienza'', I-00185, Roma, Italy.
}
\address{
  ${}^\S$\
  ICRANet and ICRA, ({\it International Center of Relativistic
Astrophysics Networks}), Piazzale della Repubblica 10, I-65122,
Pescara, Italy.}

\begin{abstract}
The cosmological constant  $\Lambda$ can be achieved as the result
of  entangled and statistically correlated  minisuperspace
cosmological states, built up by using a minimal choice of
observable quantities, i.e. $\Omega_{m}$ and $\Omega_{k}$,  which
assign the cosmic  dynamics. In particular, we consider a
cosmological model where  two regions, corresponding to two
correlated eras,  are involved; the present universe description
would be, in this way, given by a density matrix $\hat \rho$,
corresponding to an entangled final state. Starting from this
assumption, it is possible to infer some considerations on the
cosmic thermodynamics by evaluating the Von Neumann entropy. The
correlation between different regions by the entanglement
phenomenon  results in the existence of $\Lambda$ (in particular
$\Omega_{\Lambda}$) which could be interpreted in the framework of
the recent astrophysical observations. As a byproduct, this
approach could provide a natural way to solve the so called
coincidence problem.
\end{abstract}

\pacs{98.80.-k, 98.80.Jk, 98.80.Cq, 98.80.Es}

\date{\today}

\maketitle

\section{Introduction}

The  quantization of gravity is still an open question which seems
far to be achieved from several viewpoints \cite{proballa}. For
example, considering the canonical approach, the  normalization and
interpretation of the wave function, the meaning of time \cite{SS}
are puzzles difficult to be framed in coherent and self-consistent
schemes. These problems are due, directly, to the interpretation and
the solution of the  Wheeler-De Witt equation \cite{Hartle:1986gn},
and, more generally, to the  way in which one quantizes space-time.

Some theoretical efforts have been proposed  to solve these
difficulties, see for example \cite{primiero}, but it seems that
we are far from a definitive solution. Hence,  a complete quantum
description of General Relativity (GR) could be not possible.

On the other hand, several issues remain unsolved in modern
cosmology, such as the problems of {\emph dark matter} and {\emph
dark energy} \cite{revallata}. Moreover, it seems that there are a
lot of ways  to interpret  the cosmological constant $\Lambda$
which has several implications in the interpretation of
observations \cite{tuturru}.

Quantum cosmology, instead, has the final goal to describe the
universe itself by methods and results   of quantum theory. This
implies that the universe as a whole should have a quantum
description resulting in the "classical" observed cosmological
parameters.

This fact can be viewed  as a sort of entanglement with
macroscopic degrees of freedom that  leads to the classical {\em
appearance} of macroscopic systems (in this case, the whole
observed universe), a process known as de-coherence \cite{deco}.

Hence a link between quantum information and quantum cosmology can
be easily found  considering the role of gravity. Indeed, quantum
cosmology, being nothing else but an approximation,  cannot be
considered the full quantum gravity but it can assume  a key role
to describe the observed state of the universe in the light of
quantum information. This will be the main issue on which this
paper is focused. In particular this point of view would allow us
to explain, in the context of de-coherence and entanglement, some
open questions of modern  cosmology as the cosmological constant
problem.

While quantum cosmology and quantum gravity have many
interpretative problems, indeed, quantum information theory has
recently increased its theoretical self-consistency producing
several interesting results. The most important one  has been  the
achievement that entanglement phenomena of quantum states have
been framed in robust theoretical schemes and verified through
several experimental tests (see for example
\cite{guehne,elk,carteret,horodecki,horodecki2,walborn,harald,vidal,bell,chsh}).

The idea of entanglement starts from the apparent conflict between
the superposition principle and the non-separability of the
related quantum states. It appears when  a state of two or more
subsystems of a compound quantum system cannot be factorized into
pure local states of the subsystems too. This is equivalent to say
that a state, which describes a particular physical system, could
be thought as due to two, or more than two sub-states, derived
from subsystems, which are connected between them by a
non-factorizing property \cite{libroentanglato}. To fix the ideas,
if we take into account a state like
\begin{equation}\label{statusaaa}
|\Psi\rangle=|\Psi^{(1)}\rangle|\Psi^{(2)}\rangle,
\end{equation}
this is well-factorized into two sub-states, i.e.
$|\Psi^{(1)}\rangle$ and $|\Psi^{(2)}\rangle$; consequently, the
state is not entangled, but a state like
\begin{equation}\label{statusaaa2}
|\Psi\rangle=\sum_{n=0}^{M}|\Psi^{(1)}_{n}\rangle|\Psi^{(2)}_{n}\rangle,
\end{equation}
is an entangled one. This suggests  that a good way to study the
complete state is a description by the  density matrix $\hat\rho$
where one has to prepare the state $|\Psi_1\rangle$ with
probability $p_1$, the state $|\Psi_2\rangle$ with probability
$p_2$ and so on for all the sub-states of the whole system.

In order to understand the phenomenology of the entanglement, let
us  consider how it can be realized. When two physical systems
come into interaction, some correlation of  quantum nature is
generated between them and  persists  also if any apparent
interactions is not present  and  the two systems are spatially
far \cite{73}. Let us suppose that one has a bipartite (or
multipartite) quantum state \cite{111,13,188}: The
non-separability or non-factorizability is necessary to take into
account correlations \cite{19,58,96} which can evolve with time.

Several different criteria are possible to characterize
entanglement but all of them are essentially based on equivalent
forms of non-locality in pure quantum states\footnote{These
equivalences fade when we deal with mixed states.}

Hence,  entanglement represents a  fundamental physical resource
\cite{entycomerisorsa} and it has become comparable with the ideas
of energy or entropy in  the information theory of  systems.

Due to these motivations, the  so called  entanglement-degree  has
become an important feature (see for example \cite{miochuang,111})
to quantify how much a quantum state is mixed or pure
\cite{112,13,101,177}. In this paper, we are going to show that
the thermodynamics of the universe and then the emergence of the
cosmological constant $\Lambda$ are both related to the
entanglement-degree.

The paper is organized as follows. Sect.II  is  a brief summary of
the entanglement theory.  Sect.III is devoted to the definition of
the pure and mixed  cosmological states. We show that the
evaluation of the Shannon-Von Neumann entropy $S$
\cite{termodinamicadocaiser} can be directly related to the
emergence of a quantity which can be naturally considered the
cosmological constant $\Lambda$.  Conclusions are drawn in
Sect.IV.

\section{ Entanglement and quantum cosmology}

In order to develop considerations on the role of the entanglement
in quantum cosmology, let us start with taking into account an
$N$-dimensional Hilbert space where the probabilities are $p_k =
\frac{1}{N}$, $\forall k$. The interpretation is that a given
state is maximally mixed, or physically speaking, it means that
our ignorance about the configuration is total.

The opposite situation, in other words the case in which only one of
the $p_k$ is different from zero, happens when we have pure quantum
states. The first step to quantify this situation is the
introduction of a density matrix $\hat\rho$ associated to the
quantum states. For a generic pure state $|\Psi\rangle$, the density
matrix is the projector
\begin{equation}\label{matriosca}
\hat\rho_p=|\Psi\rangle\langle\Psi|\,.
\end{equation}
The main properties of $\hat \rho$ are  easily given:
\begin{enumerate}
  \item $Tr\hat\rho=1$,
  \item $Tr\hat\rho^2\leq1$,
  \item $\langle\chi|\hat\rho|\chi\rangle\geq 0$.
\end{enumerate}

It is worth noticing that $Tr\rho^2=1$ holds only if the state is
pure, viceversa $Tr\rho^2<1$. For our aims, we will consider not
pure states and then $\hat \rho$ becomes the sum on projectors,
balanced by weights, which  represents probabilities for mixed
states \cite{miochuang}.

As a consequence, we define the \emph{purity} of a state the
quantity
\begin{equation}\label{mnmnnnbnbnn}
\mu\left[\hat\rho\right]=Tr\hat\rho^2,
\end{equation}
and then  the basic concept of \emph{linear} \emph{entropy}
\begin{equation}\label{bnbnbnbnbnbn}
S_L=\frac{N}{N-1}\left(1-\mu\left[\hat\rho\right]\right),
\end{equation}
which are two prototypes of entanglement measurements.

In particular, eq. ($\ref{bnbnbnbnbnbn}$) can be interpreted as a
first-order approximation of the  concept of entropy in the
information theory, the so-called {\emph Von Neumann entropy},
which usually describes how large is the lack of information of a
quantum state \cite{ientrusp}.  We are interested to the thermal
entanglement properties of the universe. In agreement with the
statistical picture, we need a measure of the entropy of the
system. The process of entropy measurement is related to the
concept of the so called Shannon-Von Neumann entropy
\begin{equation}
\label{enggg}
S=-Tr\left(\hat{\rho}\ln\hat{\rho}\right)\,=\,-\sum_k \lambda_k\ln
\lambda_k,
\end{equation}
which satisfies concavity, subadditivity and Araki-Lieb inequality
(see for details \cite{miochuang}). The arbitrary choice of states
on which  the trace of the above equation can be evaluated, allows
us to consider the eigenvalues  $\lambda_k$ in eq.(\ref{enggg}).

In this picture, the idea of entangled states in quantum cosmology
 assume an important role. We can take into account a cosmological model where
all the eras are linked together by entanglement and then describe
the thermodynamics of such  entangled states.

This picture can be achieved  starting from the consideration that
the entropy $S$ of a system is a macroscopic quantity which can be
determined without knowing in details the  constituent microscopic
states. In other words, a measure of entanglement is a "coarse
grained" approach capable of contributing to the global
description of the whole system.

The well-known problem of defining different vacuum states in
quantum field theory could be overcome considering $\Lambda$ as
the measure of the entanglement-degree related to the
thermodynamical state of the universe. In other words,
$\Omega_\Lambda$ at a given era  can be considered as the result
of the correlation between  cosmological "quantum" states without
knowing in details the  pure states which contribute to the
correlation itself.

\section{The cosmological states and the problem of $\Lambda$}

Let us take into account a set of observable quantities defining a
cosmological state written as a vector
\begin{widetext}
\begin{eqnarray}\label{setto}
\Upsilon\,=\,\left\{H(z), a(z), q(z), j(z), s(z), l(z),
\Omega_{m}(z), \Omega_{k}(z), \Omega_{\Lambda}(z),
T(z),\overrightarrow{o}(z)\right\},
\end{eqnarray}
\end{widetext}
where  $H(z)$ is the Hubble parameter \cite{peebles},
$q(z),j(z),s(z)$ are the  cosmographic parameters, respectively the
{\it deceleration}, the {\it jerk} and the {\it snap}
\cite{HP,statisticamea,star1,star2}, $\Omega_{m}$ is the matter
density, $\Omega_{k}$ is spatial curvature density
\cite{miowheeler}, $T$ is the temperature of CMB
\cite{termodinamicadocaiser} and $\overrightarrow{o}$ is a subset of
observable quantities which contribute to better determine the
cosmological state. All the parameters are assumed as function of
the redshift $z$ and univocally assign a cosmological state. We are
simply assuming a minisuperspace model where cosmological principle
holds and "evolution" means correlation among states. In some sense,
we are defining  cosmographies at  given eras (the redshift $z$) and
we are assuming that they are correlated without specifying a priori
a field theory. It means that no theoretical model is involved when
we, for example, measure $q(z)$; in fact, following this argument,
we note that the experimental framework is the set of Supernovae
data \cite{ciaociaobelliebrutti}, or, again, $\Omega_m$ is measured
by the abundance of matter in galaxies and so on for the other
quantities in the $\Upsilon$ set \cite{peebles}. In the vector
$\Upsilon$, any parameters capable of describing the universe
dynamics and thermodynamics can be considered. However the more the
number of parameters increases, the more the cosmic state is better
assigned but the more the phase-space increases in the number of
dimensions. At this point the question  is  how to build up a
description of universe, considering this set, in terms of quantum
states.  In other words, we are wondering if it is possible to
achieve a quantum description of universe, starting from the quantum
theory, without assigning a priori field theory like GR.  In this
way, we do not need any  quantization of gravity
\cite{adm59,adm60,adm62,GARATTINI,miowheeler}, because the framework
of GR is based  on a quantum field theory; we require, so,  no field
quantization but only a quantum description of the universe, as a
vector set, derived from ($\ref{setto}$).

It is only necessary  to consider a set of astrophysical
observables  to infer the universe dynamics which could be
achieved by entanglement measures. This  \emph{minimal choice} of
observable quantities can be used to  build up a complete
phase-space evolution representing the cosmic dynamics.  Moreover,
the sense of ''minimal choice'' is that, in order to get a
complete description of the state at a given  redshift $z$, it is
necessary to define  quantities which are not linked, or
parameterized among them.  This allows, in principle,   to achieve
a description of the universe where  variables are independent
among them.

Because of $z$-dependence of all the parameters in the vector
($\ref{setto}$), $z$ becomes not directly necessary in the set;
moreover, assuming that any coarse-grained cosmological dynamics
could be represented by a Friedmann-Roberson-Walker metric (FRW),
the dimensionality of the vector can be reduced to two\footnote{It
is easy to understand, considering the homogeneity and isotropy of
the observed universe, that to describe cosmic dynamics, only two
quantities are needed, for example $H$ and $\Omega_m$; the others,
as for example the remaining cosmographic parameters, can be
expressed in terms of them.}. A minimal choice is then
\begin{equation}\label{stati}
|\phi_{i}\rangle\equiv \left(\begin{array}{c}
\Omega_{mi}\\
\Omega_{ki}
\end{array}
\right),
\end{equation}
where we do not consider yet the $z$ dependence but we labelled
the quantities by $i=1,2$, which means that cosmological  state is
assigned by   dynamical and thermodynamical parameters connecting
two given regions (i.e. epochs). The way in which we link the two
regions between them is the entanglement; in other words,  we
imagine that the epoches are correlated between them by
entanglement. A mixing hypothesis appears now as another
fundamental feature: All the regions are characterized by the fact
that they are mixed with all the others; this will be clear as
soon as  we will take into account the density matrix description
for universe states.

Let us consider now two regions and, hereafter, we will refer to
the \emph{entangled states ansatz} (ESA); this hypothesis makes us
able to build up a complete state of universe, defined first as a
mixed one only and then, as an entangled one, from eq.
($\ref{stati}$). This means that a physical state of the universe
may derive from an entanglement phenomenon between states at
different eras.

In order to construct a superposition of states, we must require
that, for $i=1,2$, the "good"  states at given eras are to be
written in form of a basis in order to describe a complete
entangled state. Following the Gram-Schmidt construction of
orthogonal and normalized basis, we have
\begin{equation}\label{stati1ver}
|e_{1}\rangle= N_{1}\left(\begin{array}{c}
\Omega_{m1}\\
\Omega_{k1}
\end{array}
\right),
\end{equation}
which is the first unitary vector;  the other one is built up by
considering the rule, (without normalization)
\begin{equation}\label{stati2ver}
|\tilde{e}_{2}\rangle=\left(\begin{array}{c}
\Omega_{m2}\\
\Omega_{k2}
\end{array}
\right)-N_{1}^{2} \left(\begin{array}{cc} \Omega_{m1} & \Omega_{k1}
\end{array}\right)\cdot
\left(\begin{array}{c}
\Omega_{m2}\\
\Omega_{k2}
\end{array}
\right) \left(\begin{array}{c}
\Omega_{m1}\\
\Omega_{k1}
\end{array}
\right),
\end{equation}
or equivalently, here with normalization
\begin{displaymath}
|e_2\rangle=N_2\left(
\begin{array}{c}
\Omega_{m2}-N_{1}^{2}\left(\Omega_{m1}^{2}\Omega_{m2}+\Omega_{m1}\Omega_{k1}\Omega_{k2}\right)\\
\\
\Omega_{k2}-N_{1}^{2}\left(\Omega_{k1}^{2}\Omega_{k2}+\Omega_{m1}\Omega_{k1}\Omega_{m2}\right)
\end{array}
\right),
\end{displaymath}
where the normalization parameters $N_1$ and $N_2$ are given by
\begin{widetext}
\begin{eqnarray}
N_1\,\equiv\,\frac{1}{\sqrt{\Omega_{m1}^{2}+\Omega_{k1}^{2}}},\,\,\,\,\,\,\,\,\,\,\,\,\,\,\,\,\,\,\,\,\,\,\,\,\,\,\,\,\,\,\,\,\,\,\,\,\,\,\,\,\,\,\,\,\,\,\,\,\,\,\,\,\,\,\,\,\,\,\,\,\,\,\,\,\,\,\,\,\,\,\,\,\,\,\,\,\,\,\,\,\,\,\,\,\,\,\,\,\,\,\,\,\,\,\,\,\,\,\,\,\,\,\,\,\,\,\,\,\,\,\,\,\,\,\,\,\,\,\,\,\,\,\,\,\,\,\,\,\,\,\,\,\,\,\,\,\,\,\,\,\,\,\,\,\,\,\,\,\,\,\,\,\,\,\,\,\,\,\,\,\,\,\,\,\,\,\,\,\,\,\,\,\,\,\,\,\,\,\,
\\
N_2\,\equiv\,\frac{1}{\sqrt{\left\{\Omega_{m2}-N_{1}^{2}\left(\Omega_{m1}^{2}\Omega_{m2}+\Omega_{m1}\Omega_{k1}\Omega_{k2}\right)\right\}^{2}+\left\{\Omega_{k2}-N_{1}^{2}\left(\Omega_{k1}^{2}\Omega_{k2}+\Omega_{m1}\Omega_{k1}\Omega_{m2}\right)\right\}^{2}}}.
\end{eqnarray}
\end{widetext}
Starting from this construction, one is able  to have entanglement
between the set of basis, postulating an entangled state, similar
to the construction of a Bell state \cite{bell}.

\subsection{The entangled states ansatz}

The ESA suggests that a possible state for describing the universe
 is
\begin{equation}\label{signdivin}
|\Psi_{\pm}\rangle\,=\,\alpha|e_1\,e_1\rangle\pm\beta|e_2e_2\rangle,
\end{equation}
with the direct expression for $|e_i e_i\rangle$
\begin{eqnarray}\label{mkooo}
|e_i\,e_i\rangle=N_{i}^{2}\left(\begin{array}{c}
\Omega_{mi}^{2}\\
\Omega_{mi}\Omega_{ki}\\
\Omega_{mi}\Omega_{ki}\\
\Omega_{ki}^{2}
\end{array}
\right),
\end{eqnarray}
and the normalization given by
\begin{equation}\label{normalyyy}
|\alpha|^2+|\beta|^2=1,
\end{equation}
in a particular Hilbert space, $\mathcal{H}$ with
$dim\mathcal{H}=4$ (such a number is not related to the number of
physical dimensions but with the minimal parameters to "assign" a
FRW universe in two eras. This is true, in terms of the expression
($\ref{mkooo}$), if we define the two {\emph simplest positions}
$\Omega_{m2}^{*}=\Omega_{m2}-N_{1}^{2}\left(\Omega_{m1}^{2}\Omega_{m2}+\Omega_{m1}\Omega_{k1}\Omega_{k2}\right)\rightarrow\Omega_{m2}$
and
$\Omega_{k2}^{*}=\Omega_{k2}-N_{1}^{2}\left(\Omega_{k1}^{2}\Omega_{k2}+\Omega_{m1}\Omega_{k1}\Omega_{m2}\right)\rightarrow\Omega_{k2}$.
Hereafter, in order to derive simple expressions for the next
calculations, we use notation in ($\ref{mkooo}$); the state of
that expression is coincident with the product of two states of
the form of eq. ($\ref{stati1ver}$) for $i=1$, but for $i=2$ it
appears possible, only if the above {\emph simplest positions}
hold.

Note that $\Omega_{m2}$ and $\Omega_{k2}$ are functions of
$\Omega_{m1}$, $\Omega_{k1}$, $\Omega_{m2}$ and $\Omega_{k2}$,
differently from $\Omega_{m1}$, $\Omega_{k1}$, which are
independent variables deduced, in principle, from observations at
given eras.

A maximally entanglement phenomenon is present if
$\alpha=\beta=\frac{1}{\sqrt{2}}$ up to a phase factor. Note that
in our picture, it is not necessary to stress that $|e_i
e_i\rangle=|e_{iA} \rangle|e_{iB}\rangle$, as mixed hypothesis
suggests, because at the same era we may imagine that
$\Omega_{\chi kA}=\Omega_{\chi kB}$, with $\chi=m,k$ and for
$k=1,2$, to simplify the model.

Considering the density matrix $\hat\rho=
\sum_{j=\pm}p_j|\phi_j\rangle\langle\phi_j|$, we have
$\hat\rho=|\alpha|^{2}|e_1e_1\rangle\langle e_1
e_1|\left(p_++p_-\right)+2\left(\alpha\beta^{*}+\alpha^{*}\beta\right)\left(p_+-p_-\right)|e_1e_1\rangle\langle
e_2 e_2|+|\beta|^{2}|e_2e_2\rangle\langle e_2
e_2|\left(p_++p_-\right)=|\alpha|^{2}|e_1e_1\rangle\langle e_1
e_1|+|\beta|^{2}|e_2e_2\rangle\langle e_2
e_2|+\left(\alpha\beta^{*}+\alpha^{*}\beta\right)\left(p_+-p_-\right)|e_1e_1\rangle\langle
e_2 e_2|$. Hence, after straightforward calculations, we get
\begin{widetext}
\begin{eqnarray}
\hat \rho\equiv |\alpha|^2N_{1}^{2}\left(
  \begin{array}{cccc}
\Omega_{m1}^{4}&\Omega_{m1}^{3}\Omega_{k1}&\Omega_{m1}^{3}\Omega_{k1}&\Omega_{m1}^{2}\Omega_{k1}^{2}\\
\\
\Omega_{m1}^{3}\Omega_{k1}&\Omega_{m1}^{2}\Omega_{k1}^{2}&\Omega_{m1}^{2}\Omega_{k1}^{2}&\Omega_{m1}\Omega_{k1}^{3}\\
\\
\Omega_{m1}^{3}\Omega_{k1}&\Omega_{m1}^{2}\Omega_{k1}^{2}&\Omega_{m1}^{2}\Omega_{k1}^{2}&\Omega_{m1}\Omega_{k1}^{3}\\
\\
\Omega_{k1}^{2}\Omega_{m1}^{2}&\Omega_{k1}^{3}\Omega_{m1}&\Omega_{k1}^{3}\Omega_{m1}&\Omega_{k1}^{4}\\
  \end{array}
\right)+\\
\\
+\left(\alpha\beta^{*}+\beta\alpha^{*}\right)\left(p_{+}-p_{-}\right)N_{1}N_{2}\left(
  \begin{array}{cccc}
\Omega_{m1}^{2}\Omega_{m2}^{2}&\Omega_{m1}^{2}\Omega_{m2}\Omega_{k2}&\Omega_{m1}^{2}\Omega_{m2}\Omega_{k2}&\Omega_{m1}^{2}\Omega_{k2}^{2}\\
\\
\Omega_{m1}\Omega_{k1}\Omega_{m2}^{2} &  \Omega_{m1}\Omega_{k1}\Omega_{m2}\Omega_{k2} & \Omega_{m1}\Omega_{k1}\Omega_{m2}\Omega_{k2} & \Omega_{m1}\Omega_{k1}\Omega_{k2}^{2}\\
\\
\Omega_{m1}\Omega_{k1}\Omega_{m2}^{2}&\Omega_{m1}\Omega_{k1}\Omega_{m2}\Omega_{k2}&\Omega_{m1}\Omega_{k1}\Omega_{m2}\Omega_{k2}&\Omega_{m1}\Omega_{k1}\Omega_{k2}^{2}\\
\\
\Omega_{k1}^{2}\Omega_{m2}^{2}&\Omega_{k1}^{2}\Omega_{m2}\Omega_{k2}&\Omega_{k1}^{2}\Omega_{m2}\Omega_{k2}&\Omega_{k1}^{2}\Omega_{k2}^{2}\\
  \end{array}
\right)+\\
\\
+|\beta|^2N_{2}^{2}\left(
  \begin{array}{cccc}
\Omega_{m2}^{4}&\Omega_{m2}^{3}\Omega_{k2}&\Omega_{m2}^{3}\Omega_{k2}&\Omega_{m2}^{2}\Omega_{k2}^{2}\\
\\
\Omega_{m2}^{3}\Omega_{k2}&\Omega_{m2}^{2}\Omega_{k2}^{2}&\Omega_{m2}^{2}\Omega_{k2}^{2}&\Omega_{m2}\Omega_{k2}^{3}\\
\\
\Omega_{m2}^{3}\Omega_{k2}&\Omega_{m2}^{2}\Omega_{k2}^{2}&\Omega_{m2}^{2}\Omega_{k2}^{2}&\Omega_{m2}\Omega_{k2}^{3}\\
\\
\Omega_{k2}^{2}\Omega_{m2}^{2}&\Omega_{k2}^{3}\Omega_{m2}&\Omega_{k2}^{3}\Omega_{m2}&\Omega_{k2}^{4}\\
  \end{array}
\right)\,,
\end{eqnarray}
\end{widetext}
where the normalized probability condition
\begin{equation}\label{prob}
p_{+}+p_{-}\,=\,1,
\end{equation}
has been used. From a physical point of view, it represents a
constraint on the  final state of the universe relating two eras.
In fact, $p_{+}$ is the probability that, the mixture representing
the universe entangled  state is in the configuration
$|\Psi_{+}\rangle\langle\Psi_{+}|$ and, on the other hand, $P_{-}$
is the probability that, the universe mixture entangled  state, is
in the configuration $|\Psi_{-}\rangle\langle\Psi_{-}|$. This is
compatible with probability normalization and with ESA. As
reported in the Introduction,  $Tr\hat\rho=1$ is an invariant
expression that can be used with other constraints to have a
complete picture of the quantum nature of universe. For maximally
entangled states and real $\alpha$, $\beta$
\begin{widetext}
\begin{eqnarray}\label{trace}
\frac{t_{1}\left(\Omega_{m2}^{2}\Omega_{m1}^{2}+\Omega_{k1}^{2}\Omega_{k2}^{2}\right)+t_{2}\Omega_{m1}\Omega_{m2}\Omega_{k1}\Omega_{k2}+\Omega_{k2}^{2}\left(\Omega_{m1}^{2}+\Omega_{k1}^{2}\right)}{2\left(\Omega_{m1}^{2}+\Omega_{k1}^{2}\right)\left(\Omega_{m2}^{2}+\Omega_{k2}^{2}\right)}=1,
\end{eqnarray}
\end{widetext}
where $t_{1}=4\alpha\beta-8\alpha\beta p_{-}+1$ and
$t_{2}=-8\alpha\beta(1-2p_-)$ with the  conditions that $\alpha$
and $\beta$ are real and the states are maximally entangled.

In addition, it is  important to note that, because of
$\Omega_{mh}+\Omega_{kh}<1$ coming from the astrophysical
observations at all the epoches (i.e. $\forall h=1,2$), we can
reduce to a constraint such inequality on $\Omega_{mi,ki}$, that
is
\begin{equation}\label{sistemaofdown}
\left\{\begin{array}{c}
  \Omega_{m1}+\Omega_{k1}+\sum_{i}\Omega_{X1i}=1, \\
  \Omega_{m2}+\Omega_{k2}+\sum_{j}\Omega_{X2j}=1.
\end{array}
\right.
\end{equation}

Immediately we can imagine a straightforward interpretation of it.
Because of $\Omega_{mi}<1, \Omega_{ki}<1$,  some unknown
quantities are involved into calculations. 
These parameters, like $\Omega_{Xhk}(z)$, are functions of the redshift $z$.

If we require that the constraint is based on observable quantities, i.e.
 $\Omega_{m}$ and $\Omega_{k}$, we have no reason to think
that the $\Omega_{Xhk}(z)$ have to be different between them. This
suggests $\Omega_{Xhk}(z)=\Omega_{Xk}(z)$. The nature of it is, at
the same time, linked to the previous two densities; in this sense
it is useful to note that no dependence from $z$ is necessary,
since if $\Omega_{m}$ and $\Omega_{k}$ vary with $z$, by
conservation relations, it is not the simplest way to think that the
sum of them loses a part of mass or curvature in time, in some
unknown variables as $\Omega_{Xk}(z)$. It may be a nonsense. Hence,
from this, it is compatible with the choice of
$\Omega_{Xk}(z)=\Omega_{Xk}$ and $\Omega_{Xk}\in[0,0.3]$.

If we recall $\sum_{k}\Omega_{Xk}$ with the compact name of
$\Omega_{\Lambda}$, we conclude that: We consider a constraint which
is the fixed normalization of observable quantities energy-matter
and curvature density, given in principle, by a sum of unknown
quantities. If these quantities are summing up and defined as one
entity, it has to be considered as independent of redshift $z$
 and with no different behavior in the different eras of universe evolution (that is,
at the same time, compatible with the universe dynamics).

Because of the given meaning of $\Omega_{\Lambda}$, it is possible to
connect its existence with the lack of information of an entangled
state, described by the Von Neumann entropy, as discussed. By
construction of it, in fact, the role of eigenvalues and of entropy
is the key to understand the existence of $\Omega_{\Lambda}$
density. Note that, up to now, no cosmological constant model is
proposed, in terms of an added cosmological constant into account of
GR, because GR has not entered the ESA. This is important because
tries to understand the specific role of $\Omega_{\Lambda}$ density,
without passing through a (classical or quantum) field theory.

Each epoch, in particular, has got its own reduced density matrix,
so its own quantum description, as part of $\hat \rho$, describing
also its own dynamics. A similar matrix is defined as
\begin{equation}\label{defmatrxred}
\hat \rho^{A}=Tr_{B}\hat \rho,
\end{equation}
in fact $\hat\rho=\hat\rho^{AB}$, because is defined, in principle,
on two different spaces, $A$ and $B$, which correspond to parts of
the two eras, characterized by different $z$ evolutions. The
expression of reduced density reads\footnote{With equivalence
between $A$ and $B$.}
$\hat\rho^{A}=\hat\rho^{B}=|\alpha|^{2}|e_1\rangle\langle
e_1|+|\beta|^{2}|e_{2}\rangle\langle e_{2}|$.

The eigenvalues\footnote{Note that
$\lambda_1\ln\lambda_1=\lambda_2\ln\lambda_2=0$ because
$\lambda_{1,2}=0$ and entropy in this case is defined $0\ln0=0$.}
calculation of density matrix $\hat\rho$ allows us to write down
entropy as follows
\begin{equation}\label{k}
S\,=\,-\left(\lambda_3\ln\lambda_3+\lambda_4\ln\lambda_4\right),
\end{equation}
which is a direct consequence of the arbitrary choice of the state
in evaluating the trace. The complete forms of eigenvalues are very
difficult to write down; to fix the ideas, in the case of
$\Omega_{k1}\approx 0$, for $p_{+}=p_{-}$\footnote{We do not expect
that this condition is crucial and physically important; a mixture
of external products among states, with the same weight, allows us
only to simplify the $\hat \rho$ expression.}, they read
\begin{equation}\label{eigenvalues}
\begin{array}{c}
\lambda_1\,=\,0,\\
\,\\
\lambda_2\,=\,0,\\
\,\\
\lambda_3\,=\,\frac{1}{2}\left(1+\frac{\sqrt{2\Omega_{m2}^{2}\Omega_{k2}^{2}\left(\alpha^2-\beta^2\right)^{2}+\Omega_{m2}^{4}+\Omega_{k2}^{4}\left(\alpha^2-\beta^2\right)^2}}{\Omega_{m2}^{2}+\Omega_{k2}^{2}}\right),
\\
\\
\lambda_4\,=\,-\frac{1}{2}\left(1-\frac{\sqrt{2\Omega_{m2}^{2}\Omega_{k2}^{2}\left(\alpha^2-\beta^2\right)^{2}+\Omega_{m2}^{4}+\Omega_{k2}^{4}\left(\alpha^2-\beta^2\right)^2}}{\Omega_{m2}^{2}+\Omega_{k2}^{2}}\right).
\end{array}
\end{equation}

For maximally entangled states, we infer the expression for
$\Omega_{\Lambda}$, in the case of
$\Omega_{k1}\approx0$\footnote{For the sake of simplicity, this
result is written in the easy case of $p_-=\frac{1}{2}$ and, here,
restoring, also, the definitions of $\Omega_{m2}$ and $\Omega_{k2}$,
without the simplest positions.}
\begin{widetext}
\begin{equation}\label{OMEGALAMBDA22}
\Omega_{\Lambda}=-\frac{\left(\Omega_{m2}^{2}+\Omega_{k2}^{2}\right)\Omega_{m1}^{5}-\left(p_-\Omega_{m2}^{2}-\Omega_{m2}^{2}-\Omega_{k2}^{2}/2\right)\Omega_{m1}^{4}-2\Omega_{m1}^{3}\Omega_{m2}^{2}+2p_+\Omega_{m1}^{2}\Omega_{m2}^{2}+\Omega_{m1}\Omega_{m2}^{2}+2p_-\Omega_{m2}^{2}-\Omega_{m2}^{2}}{\left(\Omega_{m1}^{2}+\Omega_{k1}^{2}\right)\Omega_{m1}^{4}-\Omega_{m2}^{2}\left(2\Omega_{m1}^{2}-1\right)}.
\end{equation}
\end{widetext}
If the behavior of states is not maximally entangled and no
significative approximations are done, then the most general
expressions of $\Omega_\Lambda$, with minimal positions again, for
brevity, reads
\begin{widetext}
\begin{eqnarray}\label{mm}
\Omega_{\Lambda}&=&\frac{-f_1(\Omega_m)\Omega_{m1}^{3}
-f_2(\Omega_m)
\Omega_{m1}^{2}+f_3(\Omega_m)\Omega_{m1}-f_1(\Omega_m)\Omega_{k1}^{3}+
f_4(\Omega_m)\Omega_{k1}^{2}}{\left(\Omega_{m1}^{2}+\Omega_{k1}^{2}\right)\left(\Omega_{m2}^{2}+\Omega_{k2}^{2}\right)},
\\ \nonumber
\\ \nonumber
f_1(\Omega_m):&=&\Omega_{k2}^{2}+\Omega_{m2}^{2},\\
\\ \nonumber
f_2(\Omega_m):&=&2\Omega_{m2}^{2}\alpha\beta
p_-+\Omega_{k1}\Omega_{m2}^{2}-\alpha^{2}\Omega_{k2}^{2}-2\Omega^{2}\alpha\beta
p_--\Omega_{m2}^{2}\alpha^{2}-\beta^{2}\Omega_{k2}^{2}+\Omega_{k1}\Omega_{k2}^{2}-\Omega_{m2}^{2}\beta^{2},\\
\\ \nonumber
f_3(\Omega_m):&=&-\Omega_{k1}^{2}\Omega_{k2}^{2}-\Omega_{k1}^{2}\Omega_{k2}^{2}-4\Omega_{k2}\alpha\beta\Omega_{k2}
\Omega_{k1}p_-+4\Omega_{m2}\alpha\beta\Omega_{k2}\Omega_{k1}p_+,\\
\\ \nonumber
f_4(\Omega_m):&=&\beta^2\Omega_{k2}^2+\Omega_{m2}^{2}\beta^2+\alpha^2\Omega_{k2}^{2}+2\alpha\beta
p_+\Omega_{k2}^{2}+\Omega_{m2}^{2}\alpha^2-2\alpha\beta
p_-\Omega_{k2}^{2}.
\end{eqnarray}
\end{widetext}

The entanglement measurements are various and different, as
mentioned in the introduction, and so, allowing a variation of  entropy,
in terms of $\Omega_{m2}$ and $\Omega_{k2}$, we infer the grade of
entanglement, having  an idea of its variation with
parameters of the second era.

\subsection{The coincidence problem and the interpretation of $\Lambda$}

The coincidence problem is a puzzling situation which appears
in today cosmology. It is due, essentially,  to   the resulting same order of magnitude  for
$\Omega_{m}$ and $\Omega_{\Lambda}$. From the previous discussion,
 it could be addressed by means of  eq.
($\ref{mm}$) for  $\Lambda$. It is evident, in fact, that the condition
\begin{equation}\label{ajvhac}
0<\Omega_{\Lambda}<1,
\end{equation}
can be easily achieved  and so, according to the observations, this could be seen as an
explanation of the coincidence problem. Indeed, because of the
meaning of $\Omega_{\Lambda}$, induced by the entanglement,
 it is evident, for construction, that its order of
magnitude is the same of $\Omega_{m}$ and  $\Omega_{k}$.
The interpretation of $\Lambda$  is the following: 
\\ 

\noindent  \textit{The observed value
of  cosmological constant $\Lambda$  derives from
entanglement phenomena among cosmological eras. It is the result of
 an average statistical process generated by   the superposition of several cosmic quantum states. 
Astrophysical observations are measuring the result of such a superposition.}

\section{Conclusion and perspectives}

We have studied  thermal properties of
universe and interpreted  the
cosmological constant problem in view of the so-called
entangled states ansatz (ESA). Hence the idea of two 
discrete states,  corresponding  to the measurements of  
 mass and curvature density at a  given era is
presented;  such states can result entangled giving rise to the cosmological constant. 
This view is alternative to the standard GR approach where $\Lambda$ is imposed \textit{by hand} in the cosmic evolution.
In particular,  the proposed toy model would give a straightforward approach to connect the today observed (classical) universe to 
quantum cosmology.  In a sense, also the today observed universe is a quantum state resulting from various entanglement processes. 
The role of Friedmann cosmology in this approach is to
link the emerging value of  $\Omega_\Lambda$ to the  various epochs since cosmological parameters depend on  the redshift $z$. 
In this picture,  the ESA hypothesis suggests a precise role of a cosmological states, instead of standard techniques, where
geometric  and matter degrees of freedom  have to be decomposed to achieve  quantization in generic curved space-times \cite{birrel, wippy, gr94, pon89, adm59, adm60, adm62,
hel80, mb89, bingo, 38www, 117www, hh96, gov91, ef} .

By considering two cosmic eras  and imposing that they are 
correlated by a minimal choice of observable quantities, it is
possible to use the Von Neumann entropy $S$ as an
information tool to evaluate such a correlation.  Thermodynamics is recovered by assuming 
 $S\rightarrow k_B S$ which means  that the entropy of the universe is derived by quantum entangled
states. The phenomenon of entanglement enters the model, in terms of
statistical considerations \cite{statisticamea}. The result is  that the standard cosmological constant $\Lambda$
emerges  from  an average process.

Also coincidence problem is solved in this context:  the present values of
$\Omega_{m}$ and $\Omega_\Lambda$  result of the comparable order of magnitude assuming $\Lambda$ as emerging from  an entanglement process.
The next step in this program is to achieve all the observed cosmographic parameters as the  result of  entanglement.  
This could  be a  confirmation for the present approach since cosmography is, in  general,  a robust  test bed for cosmological models \cite{salzano,mariam}.
The present paper, in this sense, is the  starting point of this program.

Moreover it seems possible to connect the 
statistical interpretation of $\Lambda$ \ with a genuine quantum field theory
interpretation since  $\Lambda$ can be the eigenvalue of  general relativistic Hamiltonians \cite{GARATTINI}.  
In this way,  the ESA hypothesis could be fully framed in a canonical quantization procedure.

\end{document}